\documentclass{article}
\usepackage{graphicx}
\usepackage{cite}
\usepackage[a4paper,left=40mm, right=30mm, top=20mm, bottom=20mm]{geometry}
\newcommand{\tr}{\mathop{\textrm{tr}}}
\renewcommand{\title}[1]{\begin{center}{\LARGE\textbf{#1}}\end{center}}
\renewcommand{\author}[1]{\begin{center}{\large{#1}}\end{center}}
\newcommand{\address}[1]{\begin{center}\textit{#1}\end{center}}

\newcommand{\pacs}[1]{PACS: {#1}}

\begin{document}

\title{Spectral and transport properties of one-dimensional nanoring superlattice}

\author{M.~A.~Pyataev\footnote{pyataevma@gmail.com} and M.~A.~Kokoreva}

\address{Institute of Physics and Chemistry, Mordovian State University,\\
Bolshevistskaya 68, Saransk, 430005, Russia}

\begin{abstract}
Spectral properties of periodic one-dimensional array of nanorings in a magnetic field are investigated.
Two types of the superlattice are considered. In the first one, rings are connected by short one-dimensional wires while in the second one rings have immediate contacts between each other. The dependence of the electron energy on the quasimomentum is obtained from the Schr\"odinger equation for the Bloch wave function.
We have found an interesting feature of the system, namely, presence of discrete energy levels in the spectrum.
The levels can be located in the gaps or in the bands depending on parameters of the system.
The levels correspond to bound states and electrons occupying these levels are located on individual
rings or couples of neighboring rings and do not contribute to the charge transport.
The wave function for the bound states corresponding to the discrete levels is obtained.
Modification of electron energy spectrum with variation of system parameters is discussed.
\end{abstract}


\section{Introduction}

The electron transport in various one-dimensional serial structures
attracts considerable attention in the last few years.
In particular, transmission properties of multiply connected 
stubs,\cite{stubs}
quantum dots\cite{dots}
and quantum rings in linear\cite{ring1,ring2,ring3,ring4,Duclos08}
and rectangular arrays\cite{rectarray} 
have been studied theoretically.
The interest to the systems is stipulated by the possibility
to tune the resistance in a very wide range
by application of electric or magnetic field and the possibility to obtain
specific transport properties of the system by variation of its geometry.
The superlattice made of nanorings is of particular interest because
the quantum ring is one of the simplest systems which
exhibits quantum interference phenomena
such as Aharonov--Bohm oscillations\cite{ABoscill} 
persistent current\cite{pc} 
and Fano resonances\cite{Fano61,FR2,Geiler03,Kokoreva11}.
A number of experimental studies have shown that quantum interference effects can be observed
at individual quantum 
rings\cite{experim1}
and two-dimensional arrays\cite{experim2}
of rings.

Theoretical investigations of chains containing finite number of elements\cite{ring1,ring2}
have shown that the conductance of the system demonstrates the miniband structure of the electron energy spectrum
even at a few rings in the chain. That means the dependence of conductance on Fermi energy exhibits a series of
almost rectangular up-and-down steps corresponding to bands and gaps in the spectrum.
Finite number of elements in the structure leads to appearance of oscillations at the bounds of the steps.
The increase of this number makes the form of the conductance closer to the ideal rectangular pattern.

The most of studies of electron transport in arrays of quantum rings
are based on the scattering matrix approach. This approach is useful due to its universality.
However it does not allow to reveal some interesting features of the electron energy spectrum.
In particular, the existence of discrete energy levels\cite{Duclos08} in the band gap
is not detected by the approach
since those levels do not contribute to the electron transport.

\begin{figure}[htb]
\begin{center}\includegraphics[width=0.8\linewidth]{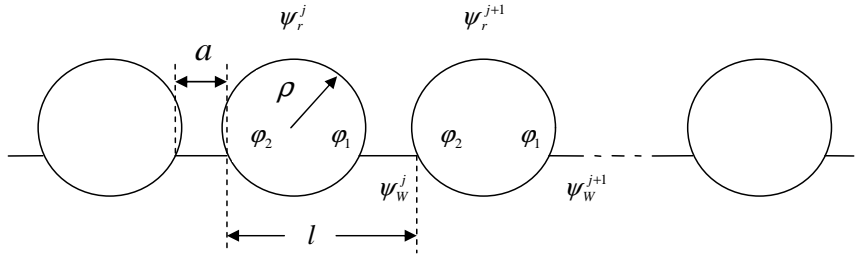}
\end{center}\caption{\label{scheme1}Scheme of the superlattice.}
\end{figure}
The purpose of the present work is the investigation of the
electron energy spectrum and the conductance of an infinite
periodic one-dimensional structure made of Aharonov-Bohm rings. We
have considered two variants of the system. The first one consists
of rings of radius $\rho$ connected by one-dimensional wires of
length $a$ (fig.~\ref{scheme1}). In the second variant, rings
have immediate point contacts between each other. In both
variants, angles defining the points of contacts are denoted by
$\varphi_1$ and $\varphi_2$. Through the paper we use polar angle
$\varphi\in[0,2\pi)$. Without loss of generality we can take
$\varphi_1=0$.

\section{Dispersion Relation}

Each point in the structure is characterized by the cell number $j$ and continuous coordinates $\xi$
in the cell, namely $\xi=\varphi$ in the ring and $\xi=x$ in the connecting wire.
We denote the electron Hamiltonian of the ring by $H_\mathrm{r}$
\begin{equation}
H_\mathrm{r}=\frac{\hbar^2}{2{m^*}\rho^2}
\left(i\frac{\partial}{\partial \varphi}+\frac{\Phi}{\Phi_0}\right)^2,
\end{equation}
where ${m^*}$ is the electron effective mass,
$\Phi=\pi \rho^2 B$ is the magnetic flux through the ring
and $\Phi_0=2\pi \hbar c/e$ is the magnetic flux quantum.
The Hamiltonian $H_\mathrm{w}$ of the electron in the wire has the form
\begin{equation}
H_\mathrm{w}=-\frac{\hbar^2}{2{m^*}}\frac{\partial^2}{\partial x^2}.
\end{equation}

In the further analysis we need the eigenvalues of the Hamiltonian $H_\mathrm{r}$
\begin{equation}
E_m=\varepsilon\left(m+\eta\right)^2,
\end{equation}
where $\varepsilon=\hbar^2/2{m^*}\rho^2$ and $\eta=\Phi/\Phi_0$.

According to the Bloch theorem the wave function should satisfy the condition
\begin{equation}
\label{psi_Bloch}
\psi^{j+1}(\xi)=e^{iql}\psi^{j}(\xi),
\end{equation}
where $\psi^{j}(\xi)$ is the wave function in $j$-th cell, $l$ is the period of the structure
and $q$ is the quasimomentum. We denote the wave functions on the ring and on the wire
by $\psi^{j}_\mathrm{r}(\varphi)$ and $\psi^{j}_\mathrm{w}(x)$ respectively.

The contact between the rings and the wires is defined
by linear boundary conditions for the wave function and its derivative.
The general form of the boundary conditions can be obtained from the
current conservation law\cite{Duclos08,Geiler03,Kokoreva11}.
We will restrict ourself by the case of continues wave function
that corresponds to equal effective width of rings and connecting wires.

The condition of continuity can be written in the following form
\begin{eqnarray}
\label{b1-1}
&&\psi^j_\mathrm{w}(0)=\psi^j_\mathrm{r}(\varphi_1),\\
\label{b1-2}
&&\psi^j_\mathrm{w}(a)=\psi^{j+1}_\mathrm{r}(\varphi_2).
\end{eqnarray}
The current conservation law requires that the sum of wave function derivatives calculated
in the outgoing direction from the contact was proportional to the value
of the wave function at the contact point
\begin{eqnarray}
\label{b2-1}
&&\psi'^{j}_\mathrm{r}(\varphi_1+0)-\psi'^{j}_\mathrm{r}(\varphi_1-0)+\rho\psi'^{j}_\mathrm{w}(0)
=v_1\psi^{j}_\mathrm{r}(\varphi_1),\\
\nonumber &&\psi'^{j+1}_\mathrm{r}(\varphi_2+0)-\psi'^{j+1}_\mathrm{r}(\varphi_2-0)-\rho\psi'^{j}_\mathrm{w}(a)\\
\label{b2-2}
&&=v_2\psi^{j+1}_\mathrm{r}(\varphi_2),
\end{eqnarray}
where $\psi'_\mathrm{w}$ is the derivative of $\psi_\mathrm{w}$ on $x$ and
$\psi'_\mathrm{r}$ is the derivative of $\psi_\mathrm{r}$ on angle $\varphi$.
As follows from Eqs.~(\ref{b2-1}) and (\ref{b2-2}) each contact is described
by one real dimensionless parameter $v_j$ that determines the strength of the point perturbation in the contact.
Positive $v_j$ corresponds to repulsive potential (barrier) in the contact
and negative value is responsible for attractive potential (well).
We note that Eqs.~(\ref{b2-1}) and (\ref{b2-2}) represent more general form of boundary conditions
than considered usually special case\cite{Xia92} of $v_j=0$ 
that corresponds to the contacts without perturbation. 
In the present paper, we will restrict ourself by the case of identical contacts $v_1=v_2=v$.

To obtain the energy spectrum of the system we construct a solution of the Schr\"odinger  equation
in each part of the elementary cell and then glue the wave function at contact
points using the Bloch theorem and boundary conditions.
It is obvious that the wave function in the wire is the superposition of propagating waves
\begin{equation}
\label{psi_w}
\psi_\mathrm{w}^j(x,E)=\alpha_1^j(E) e^{ikx}+\alpha_2^j(E) e^{-ikx},
\end{equation}
where $\alpha_1^j$ and $\alpha_2^j$ are some coefficients and $k=\sqrt{2{m^*} E}/\hbar$.

Since the contacts create point perturbations for the electron motion on  the ring
the solution of the Schr\"odinger equation can be found with the help
of the zero-range potential theory\cite{Demkov88,Albeverio88} that was effectively used
before for investigation of various nanosystems with point contacts.\cite{point-contact}
According to this theory the electron wave function on the ring for $E\neq E_m$
can be represented in terms of the Green function
$G(\varphi,\varphi_i;E)$ of the Hamiltonian $H_\mathrm{r}$
\begin{equation}
\label{psi_r1}
\psi_\mathrm{r}^j(\varphi,E)=\beta_1^j(E)G(\varphi,\varphi_1;E)+\beta_2^j(E)G(\varphi,\varphi_2;E),
\end{equation}
where $\beta_1^j$ and $\beta_2^j$ are some coefficients.
The Green function of the Hamiltonian $H_\mathrm{r}$ is well-known\cite{Geiler03}
\begin{eqnarray}
\nonumber G(\varphi,\varphi_i;E)&=&\frac{m^*}{2\hbar^2k}
\left[\frac{\exp\left(i(\varphi_i-\varphi\pm\pi)
(\eta-k\rho)\right)}{\sin{\pi(\eta-k\rho)}}\right.\\
\label{fGreen} && \left.-\frac{\exp\left(i(\varphi_i-\varphi\pm\pi)
(\eta+k\rho)\right)}{\sin{\pi(\eta+k\rho)}}\right],
\end{eqnarray}
where ``plus'' sign corresponds to $\varphi \geq \varphi_i$ and ``minus'' sign should be used otherwise.

We note that if electron energy coincides with an eigenvalue $E_m$
of the Hamiltonian $H_\mathrm{r}$ then
there is another solution of the Schr\"odinger equation.
The wave function may be represented in this case by the following equation
\begin{equation}
\label{psi_m}
\psi_\mathrm{r}^j(\varphi,E_m)=\gamma_1^j e^{im\varphi}+\gamma_2^j e^{-im\varphi}.
\end{equation}
The existence of this solution leads to appearance of an interesting phenomenon,
namely the discrete levels in the band structure.

Applying conditions (\ref{b1-1}) -- (\ref{b2-2}) to the wave
function given by Eqs.~(\ref{psi_w}) and (\ref{psi_r1}), we
obtain the system of four equations for the coefficients $\alpha_i^j$ and $\beta_i^j$ ($i=1,2$).
Existence of a nontrivial solution for this system leads to the following equation for energy
\begin{eqnarray}
\label{disperse1}
\nonumber &&2k\rho(Q_{21}e^{iql}+Q_{12}e^{-iql})-2k\rho[v\det Q+\tr Q]\cos(k a)+\\
&&\left[\det Q\left(k^2\rho^2-v^2\right)-2v\tr Q-4\right]\sin(k a)=0,
\end{eqnarray}
where $Q_{ij}(E)=\frac{\hbar^2}{m^*\rho}G(\varphi_i,\varphi_j,E)$
is the so-called Krein's Q-matrix\cite{Albeverio88,Krein47}, $\det Q$ and $\tr Q$ are the
determinant and the trace of Q-matrix respectively.
We note that Eq.~(\ref{disperse1}) is valid only for the system
with one ring in the elementary cell. In particular, this equation
describes the superlattice made of rings with diametrically opposite contacts at
arbitrary magnetic field or the system with arbitrary positions
of contacts at zero magnetic field.

\section{Rings with Immediate Contacts}

We will start our analysis from the  supperlattice constructed of the rings with immediate contacts
between each other ($a=0$). The boundary conditions in this case may be represented as follows
\begin{eqnarray}
\label{b3_1}
&&\psi^j_\mathrm{r}(\varphi_2)=\psi^{j+1}_\mathrm{r}(\varphi_1),\\
\nonumber &&\psi'^j_\mathrm{r}(\varphi_1+0)-\psi'^j_\mathrm{r}(\varphi_1-0)\\
\label{b3_2}&&+\psi'^{j+1}_\mathrm{r}(\varphi_2+0)-\psi'^{j+1}_\mathrm{r}(\varphi_2-0)
=v_s \psi^j_\mathrm{r}(\varphi_2),
\end{eqnarray}
where $v_s=v_1+v_2$.
The dispersion relation (\ref{disperse1}) acquires the form
\begin{equation}
\label{disperse2}
v_s\det Q+2\tr Q - 2(Q_{21}\emph{e}^{iql}+Q_{12}\emph{e}^{-iql})=0.
\end{equation}
We note that Eq.~(\ref{disperse2}) is valid either for the superlattice with diametrically
opposite contacts in arbitrary magnetic field or for the zig-zag superlattice with
arbitrary contact position in the zero magnetic field.
The general case of zig-zag superlattice in arbitrary magnetic field is considered in the next section.

\begin{figure}[htb]
\begin{center}\includegraphics[width=0.8\linewidth]{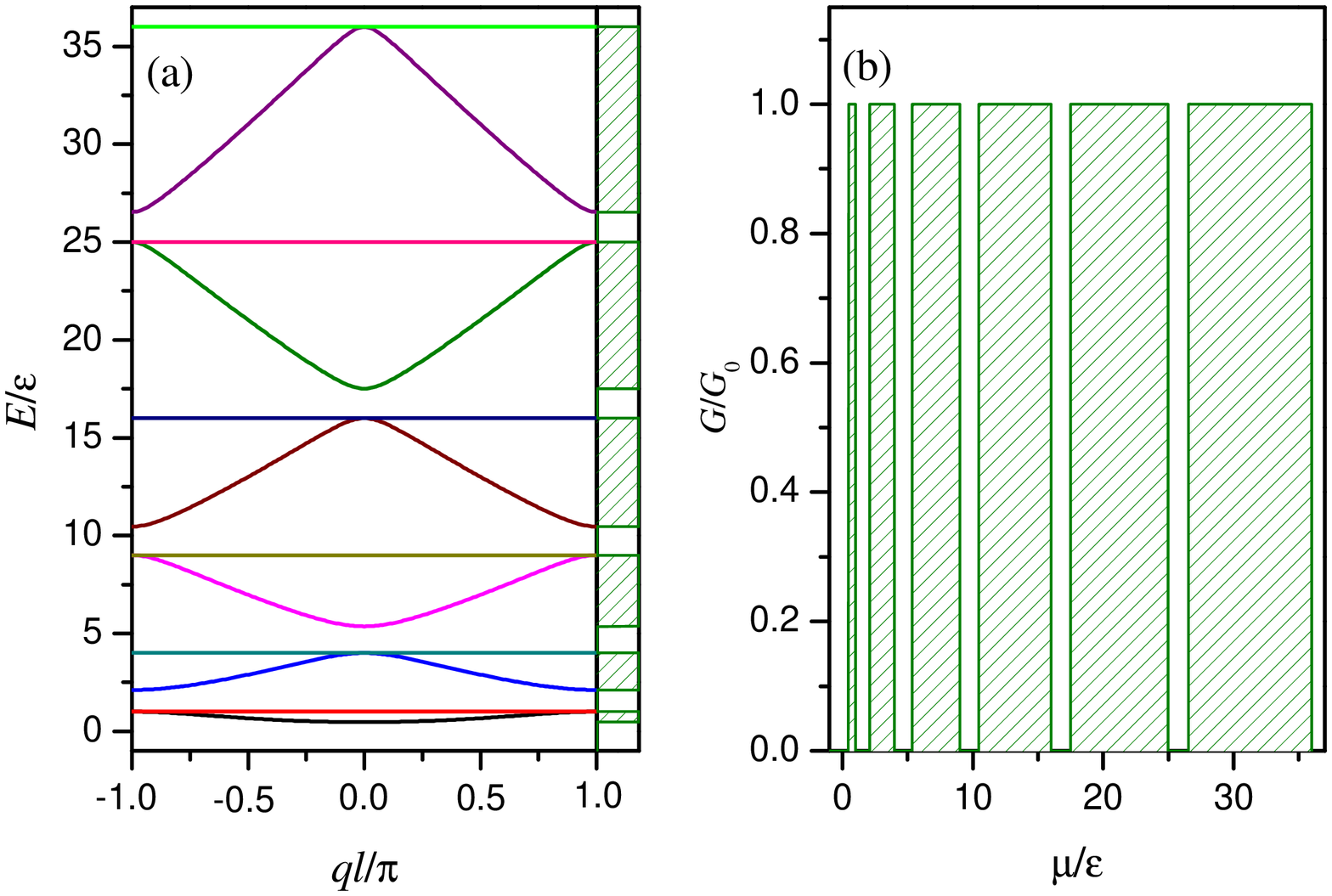}
\end{center}\caption{\label{f1withoutwire} (a) Dependence of
electron energy on quasi-momentum in the first Brillouin zone at
$a=0$, $v_s=5$, $\Phi=0$, $\varphi_2-\varphi_1=\pi$. Minibands are
shown at right side of the figure by hatched bars. (b)
Zero-temperature conductance corresponding to the spectrum (a) as
a function of chemical potential. }
\end{figure}

The dependence of the electron energy on the quasi-momentum for the simplest case of diametrically
opposite contacts ($\varphi_2=\pi$) and zero magnetic field ($\Phi=0$)
is represented in Fig.~\ref{f1withoutwire}(a). Energy bands are shown on the right hand side of the
figure by hatched bars.
We note that straight chain of nanorings without external magnetic field has
been considered earlier\cite{Duclos08} by P. Duclos and coauthors.
However, for convenience of reader, we present here the results concerning this case.
The dispersion relation (\ref{disperse2}) for this conditions may be written in the following simple form
\begin{equation}
\label{disperse2a}
v_s\sin(\pi k\rho)+4k\rho[\cos(\pi k\rho)-\cos(ql)]=0.
\end{equation}
The similar equation was obtained in Ref.~\citen{Duclos08}.

One can see that the energy spectrum has band structure without
band crossings. According to the Landauer-B\"uttiker formalism the
dependence of the zero-temperature conductance on chemical
potential replicates the electron energy band structure
(Fig.~\ref{f1withoutwire}(b)). Since there is no band crossings in the
system the conductance does not exceed one conductance quantum.

\begin{figure}[thb]
\begin{center}\includegraphics[width=0.8\linewidth]{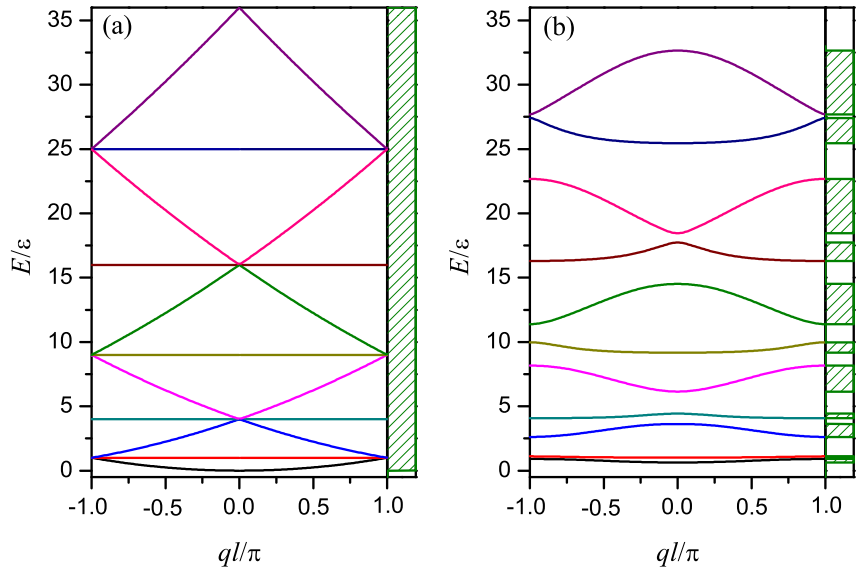}
\end{center}\caption{\label{f1withoutwire2}   Dependence of
electron energy on quasi-momentum at $a=0$, $\Phi=0$, (a) $v_s=0$,
$\varphi_2-\varphi_1=\pi$, (b) $v_s=10$,
$\varphi_2-\varphi_1=0.95\pi$.}
\end{figure}
An interesting feature of the system is presence of discrete levels at values  $E_m$ at the edge of the energy band
(Fig.~\ref{f1withoutwire}(a)).
To study the appearance of these levels we consider the spectrum of the Hamiltonian $H_\mathrm{r}$.
We note that the energy levels $E_m$ of isolated ring are double degenerated at $m\neq 0$
and the general form of the eigenfunction is given by Eq.~(\ref{psi_m}).
The degeneration degree in the set of $N$ independent rings is equal to $2N$.
Presence of contacts between the rings can remove the degeneration partially or completely.
Namely, at diametrically opposite contacts ($\varphi_2=\pi$) one
half of states is transformed into an energy band while the other half is conserved
as a discrete level at $E_m=\varepsilon m^2$.
Such levels correspond to localized electrons which are not involved in the charge transfer process.
The electron wave function for this level vanishes at both contact points
and electrons are confined on individual rings.
In contrast to the bound states in continuum stipulated by a single defect or an adatom\cite{adatom}
the states in the superlattice of quantum rings are highly degenerated
(the degeneration  degree is infinite in the strictly periodic system).

Let us assume that the electron is confined on the $n$-th ring.
Then the wave function corresponding to this state has the form
\begin{equation}
\label{discrete}
\psi^j_\mathrm{r}(\varphi)=\left\{
\begin{array}{ll}
\gamma_j\sin(m\varphi),&j=n,\\
0,&j\neq n.
\end{array}\right.
\end{equation}
Of course, any linear combination of the functions given by Eq.~(\ref{discrete}) corresponds to the same energy.
Therefore the degeneration degree of level $E_m$ is equal to the number of rings in the system.
Actually the level $E_m$ exists in the spectrum if $\varphi_2=l\pi/n$, where $n$ is an integer divisor of $m$ and $l$ is
an arbitrary integer number.
\begin{figure}[!ht]
\begin{center}\includegraphics[width=0.8\linewidth]{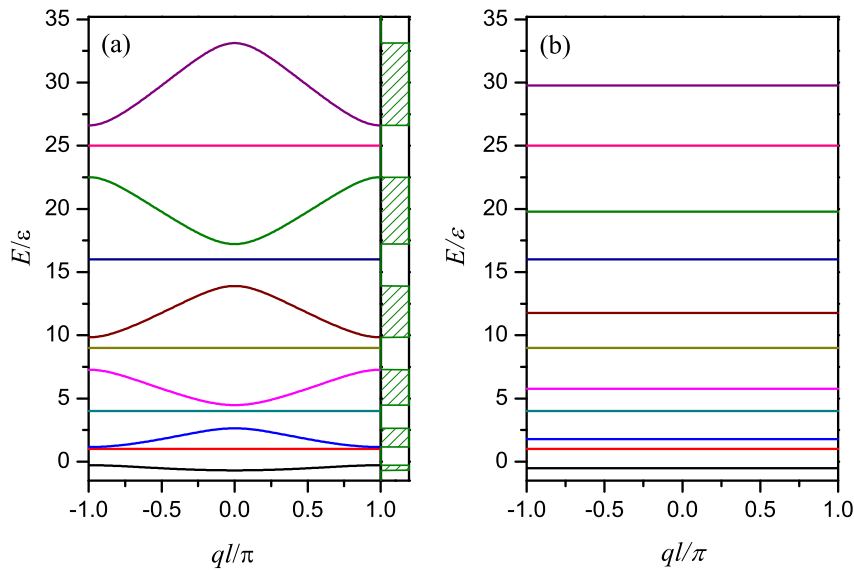}
\end{center}\caption{\label{f2withoutwire}   Dependence of
electron energy on quasi-momentum for the rings with immediate
contact at $\varphi_2-\varphi_1=\pi$, $v_s=-3$. Magnetic flux is
(a) $\Phi=0.2\Phi_0$, (b) $\Phi=0.5\Phi_0$.  All minibands
degenerate into discrete levels when the magnetic flux is a
half-integer.}
\end{figure}
It is easy to construct a linear combination of the Bloch type from the functions given by Eq.~(\ref{discrete}).
This linear combination might seam to correspond to delocalized electrons since the probability
of finding the particle on a given ring is equal for all rings. However, the electron velocity
corresponding to such a wave function is zero because the energy is independent of quasimomentum.

If $\eta =0$ and $\varphi_2 =\pi$ then the level $E_m$ is located
exactly at the edge of energy bands as it follows from Eq.~(\ref{disperse2a}).
The sequence of bands and discrete levels depends on the sign of the contact parameter $v_s$.
If $v_s<0$ (contact perturbation corresponds to the quantum well)
then the bands are located below the level $E_m$ and vice versa (Fig.~\ref{f1withoutwire}(a)).
All bands merge at $v_s=0$ and the electron energy spectrum become continuous (Fig.~\ref{f1withoutwire2}(a)).
We note that the band gaps in the spectrum disappear only if $\varphi_2=\pi$ and $\eta=0$.
In other cases the gaps are conserved in the spectrum even at ideal contacts ($v_s=0$).
Small deviation of contacts from the diametrically opposite points leads to
broadening of the discrete levels into narrow minibands (Fig.~\ref{f1withoutwire2}(b)).

The surprising result is that the level at $E_m=\varepsilon m^2$ is conserved in the magnetic field
if contacts are placed at diametrically opposite points $\varphi_1=0$ and $\varphi_2=\pi$ (Fig.~\ref{f2withoutwire}(a)).
In this case, all elements of Q-matrix vanish at $E=E_m$ and Eq.~(\ref{disperse2}) is satisfied for all $q$.
In the magnetic field, the energy bands are shifted and the discrete level $E_m$ is located in the band gap.
The wave function corresponding to this level can be found from Eqs.~(\ref{psi_r1}) and (\ref{fGreen}).
It vanishes at both contact points and satisfies the condition (\ref{b3_2}).
We note that energy $E_m=\varepsilon m^2$ is not an eigenvalue for the Hamiltonian $H_\mathrm{r}$ of isolated ring
and therefore there is no smooth functions satisfying the boundary conditions in this case.
The wave function corresponding to $E_m=\varepsilon m^2$ has a kink either at $\varphi=0$ or at $\varphi=\pi$.
An electron could not be confined on a single ring because the corresponding wave function does not satisfy
the condition (\ref{b3_2}). However the electron could be localized on a couple of rings.
Assume these rings have numbers $n$ and $n+1$ then the wave function has the form
\begin{equation}
\label{psi_nn1}
\psi^j(\varphi)=\left\{
\begin{array}{ll}
\gamma_n e^{-i\eta(\varphi\pm\pi)}\sin(m\varphi),& j=n,\\
-\gamma_n e^{-i\eta\varphi}\sin(m\varphi),& j=n+1,\\
0,& \textrm{otherwise}.
\end{array}\right.
\end{equation}
Here the ``plus'' sign corresponds to $0\leq \varphi \leq \pi$ and ``minus'' sign corresponds to $\pi<\varphi< 2\pi$.
The degeneration degree of the level $E_m=\varepsilon m^2$ in the magnetic field is still equal
to the number of cells in the structure because the states corresponding to electrons
localized at two neighbor couples are orthogonal.

We note that presence of discrete levels in the gaps should have an interesting
consequence in photoconducting properties of the system.
If the chemical potential at low temperature is located between the level and the miniband
then the conductance of the system should be increased significantly due to the radiation
induced transition from the level to the miniband.
This phenomenon could be used in production of turnable resonance photodetector.
It should be mentioned that specific property of discrete levels
in the considered superlattice is the large degeneration degree
which is equal to the number of unit cells, in contrast to the well-known impurity
levels in the band gap with the degeneration degree equal to the number of impurities.

An interesting phenomenon is found at half-integer magnetic flux
through the ring when the contacts are located at diametrically
opposite points. All minibands are collapsed into discrete levels
in this case (Fig.~\ref{f2withoutwire}b), and the conductance of the
system vanishes. If $\eta=0.5$ then $Q_{12}=0$ for all values of $E$
and therefore dispersion relation (\ref{disperse2}) gets the form
\begin{equation}
\label{eta05}
Q_{jj}(v_sQ_{jj}+4)=0.
\end{equation}
It is satisfied if either $Q_{jj}=0$ or $Q_{jj}=-4/v_s$.
Thus there are two type of levels. Levels of the first
type are given by equation $E_m=\varepsilon m^2$.
Their position is independent of $v_s$.
Levels of the second type are determined by equation
\begin{equation}
\label{eta05-2}
\tan \pi k \rho=-4k\rho/v_s.
\end{equation}
The levels of second type become closer to the levels of first type when the
coupling between rings decreases.
When there is no connection between the rings ($v_s\rightarrow \infty$) levels merge with each
other and become doubly degenerate.
It should be noted that the electron transmission coefficient of an individual
Aharonov-Bohm ring vanishes at half-integer magnetic flux\cite{Geiler03}.

\section{Zig-Zag Superlattice in a Magnetic Field}
\begin{figure}[htb]
\begin{center}\includegraphics[width=0.8\linewidth]{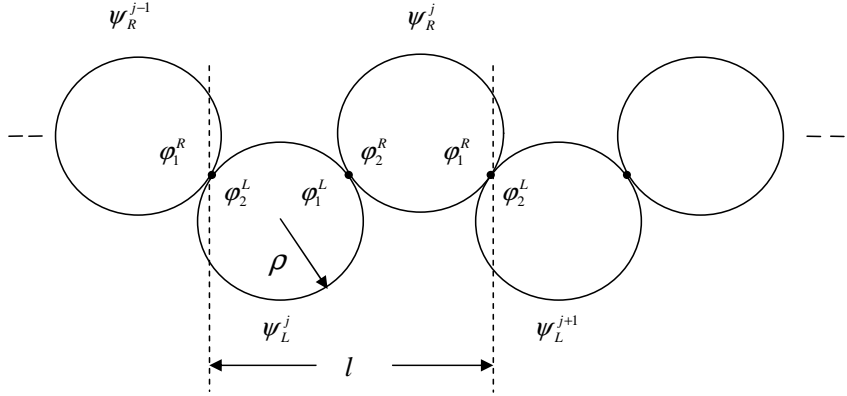}
\end{center}\caption{\label{Rings3} Schematic representation of a
Aharonov-Bohm ring superlattice with elementary cell containing
two rings.}
\end{figure}

Schematic view of the zig-zag superlattice is represented in Fig.~\ref{Rings3}.
At arbitrary magnetic field and arbitrary position of contacts two neighbor rings become inequivalent
because the system has no inversion symmetry in this case.
The unit cell of the superlattice consists of two rings. Angles determining the position of contacts are
marked by letters "$L$" and "$R$" in left and right ring respectively.
We note that angles $\varphi^L_i$ and $\varphi^R_i$ are not independent. Since the centers of rings
are located at two parallel lines, the angles satisfy the condition
\begin{equation}
\label{phi_LR}
\varphi^L_2=2\pi-\varphi^R_2.
\end{equation}
\begin{figure}[hbt]
\begin{center}\includegraphics[width=0.8\linewidth]{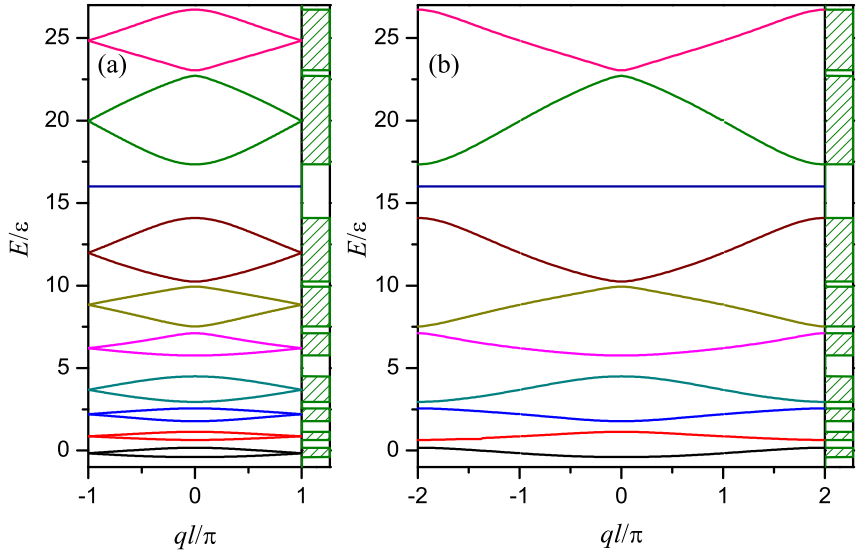}
\end{center}\caption{\label{f4}   (a) Dependence of electron
energy on quasi-momentum for zig-zag superlattice without
connected wires at $v_s=-2$,  $\varphi_{L1}=3\pi/4$,
$\varphi_{L2}=5\pi/4$, (a) $\Phi=0.2\Phi_0$, (b)  the same
dependence plotted in two Brillouin zones.}
\end{figure}
We denote by $\psi_L^j(\varphi)$ and $\psi_R^j(\varphi)$ the functions in the left and the right rings
of unit cell respectively. The boundary conditions have the form
\begin{eqnarray}
\label{bound4_1}
&&\psi_L^j(\varphi_{1}^L)=\psi_R^j(\varphi_{2}^R),\\
\label{bound4_2}
&&\psi_R^j(\varphi_{1}^R)=\psi_L^{j+1}(\varphi_{2}^L),\\
\nonumber
&&\psi'^j_L(\varphi_{1}^L+0)-\psi'^j_L(\varphi_{1}^L-0)\\
\label{bound4_3}&&+\psi'^j_R(\varphi_{2}^R+0)-\psi'^j_R(\varphi_{2}^R-0)
=v_s\psi^j_L(\varphi_{1}^L),\\
\nonumber
&&\psi'^{j+1}_L(\varphi_{2}^L+0)-\psi'^{j+1}_L(\varphi_{1}^L-0)\\
\label{bound4_4}&&+\psi'^j_R(\varphi_{1}^R+0)-\psi'^j_R(\varphi_{1}^R-0) =v_s\psi^j_R(\varphi_{1}^R).
\end{eqnarray}
Taken into account that the angles $\varphi_i^L$ and $\varphi_i^R$ are related
via condition (\ref{phi_LR}), we can write down the following relation for the Green function
\begin{equation}
\label{G_LR}
G(\varphi^L_i,\varphi^L_j;E)=G(\varphi^R_j,\varphi^R_i;E).
\end{equation}
Equation (\ref{G_LR}) allows us to use the Q-function for the only one ring (left or right) in our calculations.
Then we can represent the dispersion relation in the form
\begin{eqnarray}
\nonumber &&\left(v_s^2\det Q+2v_s\mathop{\mathrm{tr}}Q-4|Q_{12}|\cos\left(\frac{ql}{2}\right)\right)\\
\label{disperse3}&\times&\left(v_s^2\det Q+2v_s\mathop{\mathrm{tr}}Q+4|Q_{12}|\cos\left(\frac{ql}{2}\right)\right)=0.
\end{eqnarray}
The dependence of electron energy on quasimomentum in zig-zag superlattice is shown in Fig.~\ref{f4}.
If the Q-function is real then energy bands obtained from Eq.~(\ref{disperse3}) coincide
with the bands obtained from Eq.~(\ref{disperse2}).
We note that each band in Fig.~\ref{f4}(a) is formed by two branches which are crossed at $ql=\pm\pi$.
The kink on dispersion curve at $ql=\pm\pi$ can be removed if we consider the second Brillouin zone (Fig.~\ref{f4}).

One can see in Fig.~\ref{f4} the discrete level at $E=16\varepsilon$.
In agreement with the previous section the level presents in the spectrum at $\varphi_2=n\pi/4$ with integer $n$
independently on the value of magnetic field because the wave function given by Eq.~(\ref{discrete})
vanishes at both contact points.

\section{Rings Connected by Wires}

\begin{figure}[htb]
\begin{center}\includegraphics[width=0.8\linewidth]{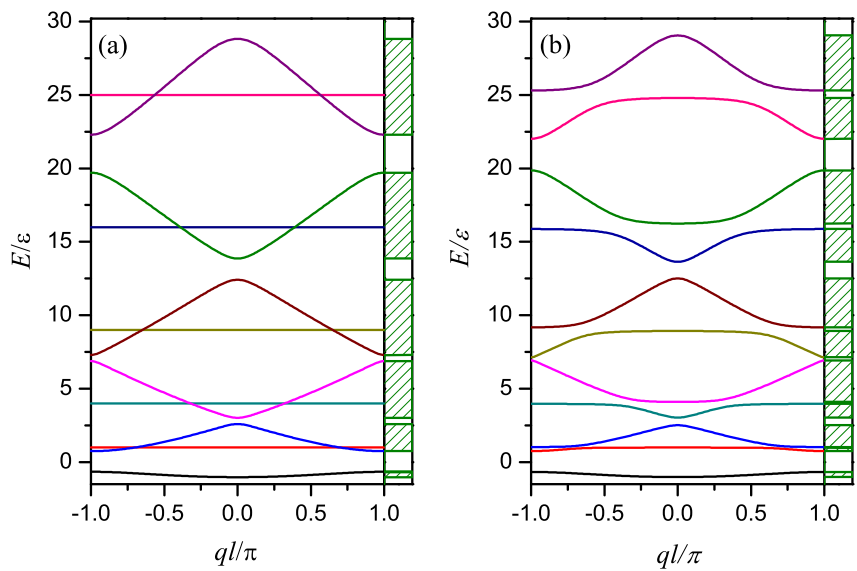}
\end{center}\caption{\label{f1withwire}  (a) Dependence of
electron energy on quasi-momentum at $\varphi_2-\varphi_1=\pi$,
$a=0.2\rho$, $v=-2$ and $\Phi=0$.  (b) The same dependence at
$\Phi=0.1\Phi_0$. }
\end{figure}

Let us consider the superlattice constructed of rings connected by wires.
Dispersion relation for the system is given by Eq.~(\ref{disperse1}).
We will start from the simplest case of diametrically opposite contacts.
The dependence of energy on quasimomentum for this case is represented in Fig.~\ref{f1withwire}.
One can see the discrete levels corresponding to localized electron.
In contrast to the case of immediate contacts some of the levels are immersed in minibands.
That means the localized and delocalized states exist at the same quasi-momentum.
The wave function corresponding to the localized states vanishes in connecting wires and has the form
of Eq.~(\ref{discrete}) in rings.
The wave function corresponding to delocalized state is obtained from Eqs.~(\ref{psi_r1}) and (\ref{fGreen})
and could be written as follows
\begin{eqnarray}
\label{psi_nn2}
\psi^j_\mathrm{r}(\varphi)=\gamma^j_1 \cos(m\varphi)\pm\gamma^j_2\sin(m\varphi),
\end{eqnarray}
where the sign is determined in the same way as in Eq.~(\ref{psi_nn1}).
In wires the wave function of delocalized states is given by Eq.~(\ref{psi_w}).
The presence of magnetic field leads to broadening of the discrete levels located in the gaps
into narrow minibands (Fig.~\ref{f1withwire}(b)).
If the band is crossed by the discrete level at zero magnetic field then
the magnetic field leads to appearing of the gap in the vicinity of the level position (Fig.~\ref{f1withwire}(b)).

\begin{figure}[htb]
\begin{center}\includegraphics[width=0.8\linewidth]{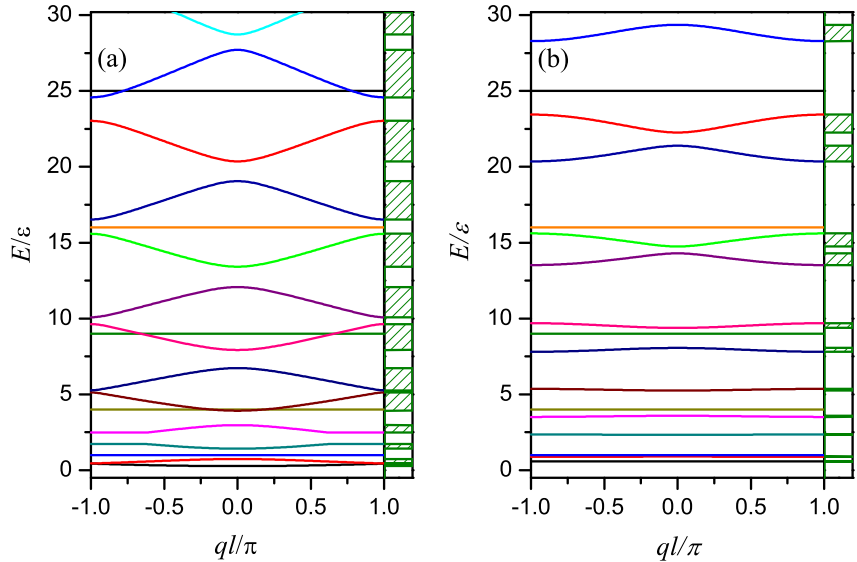}
\end{center}\caption{\label{f2}   Dependence of electron energy
on quasi-momentum at $\varphi_2-\varphi_1=\pi$, $\Phi=0$ and
$a=4\rho$, (a) $v=2$, (b) $v=20$.}
\end{figure}

With the increase of the length of connected wires the number of bands
at fixed interval of energy increases because of the enlarged
period of the structure (Fig.~\ref{f2}(a)).
The energy bands become narrower and shift to higher energies with increase in the perturbation
strength $v$ (Fig.~\ref{f2}(b)). As might be expected, the energy spectrum of the system
tends to spectrum of isolated rings in the limit $v\to\infty$.
We note that in the case of rings connected by wires energy gaps are conserved in the spectrum even at $v=0$.

\begin{figure}[htb]
\begin{center}\includegraphics[width=0.8\linewidth]{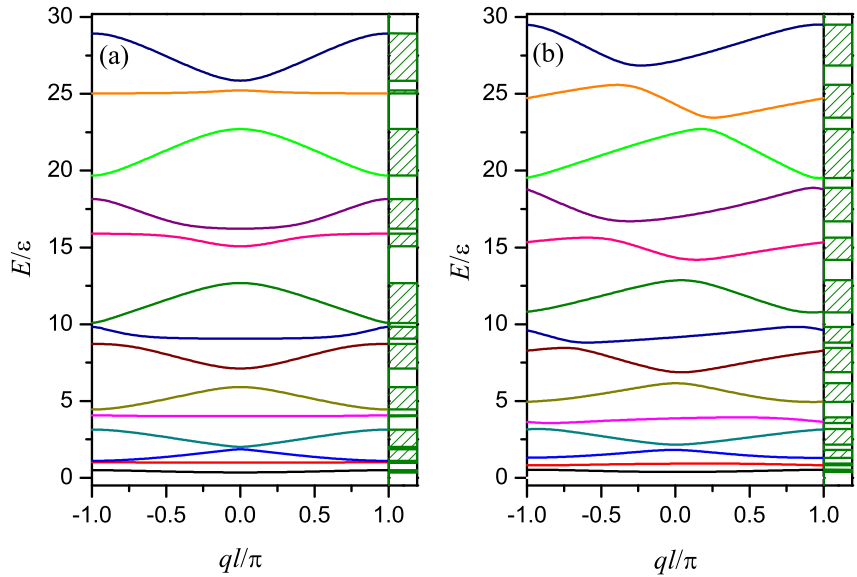}
\end{center}\caption{\label{f34}  Dependence of electron energy
on quasi-momentum at $v=2$,  $a=2\rho $,
$\varphi_2-\varphi_1=0.95\pi$, (a) $\Phi=0$; (b)
$\Phi=0.2\Phi_0$.}
\end{figure}

If the wires are connected at non-opposite points then
discrete levels are transformed into narrow minibands (Fig.~\ref{f34}(a))
similarly to the case of rings with immediate contacts.
The dependence of energy on quasi-momentum is asymmetric in the magnetic
field at non-opposite contacts because the system has no inverse symmetry in this case.
The minimum value of the conducting band and the maximum value of the valence band can be located at
different points of the Brillouin zone (Fig.~\ref{f34}(b)).
Therefore indirect miniband structure can be obtained in the system.
This band structure is interesting in relation to photogalvanic effect because
the intersubband photon absorbtion in such superlattice could be used for generation
of the photocurrent.

\section*{Conclusions}

We have studied the electron energy spectrum and zero-temperature conductance of one-dimensional super\-lattice
made of Aharo\-nov -- Bohm rings connected by one-dimensional wires.
We have found that nanorings in the magnetic field provide very wide possibilities
to construct superlattices with various miniband structures.
According to our analysis the energy spectrum has band structure without band crossings.
Consequently the conductance of the system does not exceed one conductance quantum.
The band gap width depends on the coupling between the rings, the position of contacts,
the length of connecting wire and the value of external magnetic field.
As might be expected the width of gaps increases with decrease in coupling strength between the rings
and vice versa.

An interesting result is the presence of discrete energy levels in the gap or even in the band.
These levels are originated from the double-degenerated eigenvalues of the isolated ring Hamiltonian $H_\mathrm{r}$.
We have found the conditions for appearance and disappearance of the levels. In particular,
discrete levels exist in the spectrum of superlattice if the rings have immediate contacts between each other
at diametrically opposite points.
In this case, the levels are situated at the edge of the bands and the gaps.
At non-zero magnetic field the levels are conserved in the spectrum but they are moved into the band gap.
The levels are transforms into narrow minibands if the contacts are shifted from the diametrically opposite points.
If the rings are connected by wires then the levels exist only at diametrically opposite contacts and zero magnetic
field. Small magnetic field leads to broadening of the levels into narrow bands.
The levels are degenerated with the degeneration degree equal to the number of rings in the superlattice.
The wave function corresponding to the level may be chosen as the bound state localized
on the single ring or on the couple of neighbor rings. Electrons occupying the level have zero velocity
and do not contribute to the charge transfer. In the case of diametrically opposite contacts and
half-integer magnetic flux through the ring all energy bands are degenerated into discrete levels and perfect electron
localization occurs.  If the wires are connected to the rings at non-diametrically opposite points
then the dependence of electron energy on quasimomentum becomes
asymmetric in magnetic field since the reversal symmetry of the system is broken.
This phenomenon should result in photogalvanic effect based on interminiband transitions.
We note that presence of discrete levels in the spectrum provides
interesting possibilities of application the superlattice as photodetector with turnable spectral sensitivity.

\section*{Acknowledgements}
Authors are grateful to A.V.~Shorokhov and V.A.~Margulis for valuable discussions.
The work is supported by the RFBR and the Ministry of Education and Science of Russia (project no. 11.519.11.3023).


\begin{thebibliography}{88}

\bibitem{stubs}
S. Chattopadhyay and A. Chakrabarti,
\textit{J. Phys.: Condens. Matter} \textbf{16}, 313 (2004);
%
Y.~P.~Chen, Y. E. Xie and X. H. Yan,
\textit{Phys. Rev. B} \textbf{74}, 035310 (2006);
%
P. S. Deo et al.,
\textit{Phys. Rev. B} \textbf{58}, 10784 (1998);
%
B.-Y. Gu,
\textit{Phys. Rev. B} \textbf{51}, 16840 (1995);
%
G. J. Jin et al.,
\textit{J. Appl. Phys.} \textbf{85}, 1597 (1999);
%
S. Sengupta, A. Chakrabarti and S. Chattopadhyay,
\textit{Phys. Rev. B} \textbf{71}, 134204 (2005).

\bibitem{dots}
W. Gong et al.,
\textit{Phys. Rev. B} \textbf{73}, 245329 (2006);
%
W. Gong, Y. Han and G. Wei,
\textit{J.~Phys.: Condens. Matter} \textbf{21},  175801 (2009);
%
S. Tanaka et al.,
\textit{Phys. Rev. B} \textbf{76}, 153308 (2007);
%
J. A. Brum,
\textit{Phys. Rev. B} \textbf{43},  12082 (1991).

\bibitem{ring1}
D. Takai and K. Ohta,
\textit{Phys. Rev. B} \textbf{50}, 2685 (1994);
%
D. Takai and K. Ohta,
\textit{Phys. Rev. B} \textbf{50}, 18250 (1994);
P. S. Deo and A. M. Jayannavar,
\textit{Phys. Rev. B}  \textbf{50}, 11629 (1994);
%
A. Chakrabarti, R. A. R\"omer and M. Schreiber,
\textit{Phys. Rev. B}  \textbf{68}, 195417 (2003);
%
Y.-F.~Gao and Y.-P. Zhang,
\textit{Chin. Phys. Lett.} \textbf{22}, No. 5, 1045 (2005).


\bibitem{ring2}
B. Moln\'{a}r,  P. Vasilopoulos and F. M. Peeters,
\textit{Appl. Phys. Lett.} \textbf{85}, 612 (2004);
P.~Vasilopoulos, B. Moln\'{a}r and F. M. Peeters,
\textit{Int. J. Mod. Phys. B} \textbf{18}, 3661 (2004);
%
B. Moln\'{a}r,  P. Vasilopoulos and F. M. Peeters,
\textit{Phys. Rev. B} \textbf{72}, 075330 (2005);
%
W. Cui and G. Jim,
\textit{Int. J. Mod. Phys. B} \textbf{19}, 2865 (2005).

\bibitem{ring3}
Y. Liu and P. M. Hui,
\textit{Phys. Rev. B} \textbf{57}, 12994 (1998);
%
J. Li, Z. Q. Zhang and Y. Liu,
\textit{Phys. Rev. B}  \textbf{55}, 5337 (1997).

\bibitem{ring4}
J. Yi et al.,
\textit{Phys. Rev. B} \textbf{65}, 033305 (2001);
%
W. Park and J. Hong,
\textit{Phys. Rev. B} \textbf{69},  035319 (2004);
%
D. Shin and J. Hong,
\textit{Phys. Rev. B} \textbf{72}, 113307 (2005);
%
H.-B. Xue et al.,
\textit{Chin. Phys. B} \textbf{19}, 047303 (2010).

\bibitem{Duclos08}
P. Duclos, P. Exner and O. Turek,
\textit{J. Phys. A.: Math. Theor.} \textbf{41}, 415206 (2008).

\bibitem{rectarray}
P. F\"{o}ldi et al.,
\textit{Nano Letters} {\bf 8}, 2556 (2008);
%
O. K\'{a}lm\'{a}n et al.,
\textit{Phys. Rev. B} {\bf 78}, 125306 (2008).


\bibitem{ABoscill}
Y. Aharonov and D. Bohm,
\textit{Phys. Rev.} \textbf{115}, 485 (1959);
%
M. B\"{u}ttiker, Y. Imry and  M.~Ya.~Azbel,
\textit{Phys. Rev. A} \textbf{30}, 1982 (1984);
%
Y. Gefen, Y. Imry and M.~Ya.~Azbel,
\textit{Phys. Rev. Lett.} \textbf{52}, 129 (1984);
%
A. Levy Yeyati and M. B\"{u}ttiker,
\textit{Phys. Rev. B}  {\bf 52}, R14360 (1995).

\bibitem{pc}
M. B\"{u}ttiker, Y. Imry and R. Landauer,
\textit{Phys. Lett. A} {\bf 96}, 365 (1983);
%
M. B\"{u}ttiker and C. A. Stafford,
\textit{Phys. Rev. Lett.} {\bf 76}, 495 (1996);
%
P. Cedraschi, V. V. Ponomarenko  and M. B\"{u}ttiker,
\textit{Phys. Rev. Lett.} {\bf 84}, 346 (2000).

\bibitem{Fano61}
U. Fano,
\textit{Phys. Rev. B} \textbf{104}, 1866 (1961).

\bibitem{FR2}
Y. S. Joe, A. M. Satanin and G. Klimeck,
\textit{Phys. Rev. B} \textbf{72}, 115310 (2005);
%
V. Vargiamidis and  H. M. Polatoglou,
\textit{Phys. Rev. B} \textbf{74}, 235323 (2006).

\bibitem{Geiler03}
V. A. Geiler, V. V. Demidov and V. A. Margulis,
\textit{Tech. Phys.} \textbf{48}, 661 (2003).

\bibitem{Kokoreva11}
M. A. Kokoreva, M. A. Margulis and M. A. Pyataev,
\textit{Physica E} (Amsterdam) \textbf{43}, 1610 (2011).

\bibitem{experim1}
A. Yacoby et al.,
\textit{Phys. Rev. Lett.} \textbf{74}, 4047 (1995);
%
N. T. Bagraev et al.,
\textit{Semicond.} \textbf{34}, 7,  817 (2000);
%
A. A. Bykov et al.,
\textit{JETP Lett.} \textbf{72}, 209 (2000);
%
W. G. van der Wiel et al.,
\textit{Science} \textbf{289}, 2105 (2000);
%
K. Kobayashi et al.,
\textit{Phys. Rev. Lett.}  \textbf{88}, 256806 (2002);
%
K. Kobayashi et al.,
\textit{Phys. Rev. B} \textbf{70}, 035319 (2004);
%
F. G. G. Hernandez et al.,
\textit{Phys. Rev. B} \textbf{84}, 075332 (2011).

\bibitem{experim2}
W. Rabaud et al.,
\textit{Phys. Rev. Lett.}  \textbf{86}, 3124 (2001);
%
T. Bergsten et al.,
\textit{Phys. Rev. Lett.} \textbf{97}, 196803 (2006).

\bibitem{Albeverio88}
S. Albeverio et al.,
\textit{Solvable Models in Quantum Mechanics} (Springer-Verlag, Berlin, 1988).

\bibitem{Demkov88}
Yu. N. Demkov and V. N. Ostrovsky,
\textit{Zero-Range Potentials and Their Applications in Atomic Physics} (Plenum, New York, 1988).

\bibitem{Xia92}
J. B. Xia, 
\textit{Phys. Rev. B} \textbf{45}, 3593 (1992).

\bibitem{point-contact}
V. A. Geyler and I. Yu. Popov,
\textit{Theor. Math. Phys.} \textbf{107}, 427 (1996);
%
B. S. Pavlov et al.,
\textit{Europhys. Lett.} \textbf{52}, 196 (2000);
%
J. Br\"uning and V. A. Geyler,
\textit{J. Math. Phys.} \textbf{44}, 371 (2003);
%
V. A. Geyler, V. A. Margulis and M. A. Pyataev,
\textit{JETP} \textbf{97}, 763 (2003);
%
R.~Carlone and P. Exner
\textit{Rep. Math. Phys.} \textbf{67}, 211 (2011).

\bibitem{Krein47}
M. G. Krein,
\textit{Mat. Sb.} \textbf{20}, 431 (1947).

\bibitem{adatom}
K. Nikoli\'{c} and R. \v{S}ordan,
\textit{Phys. Rev. B} \textbf{58}, 9631 (1998);
%
H. Nakamura et al.,
\textit{Phys. Rev. Lett.} \textbf{99}, 210404 (2007).

\end{thebibliography}
\end{document}